\documentstyle[12pt,preprint,eqsecnum,aps]{revtex}
\begin{document}
\begin{flushright}
MRC.PH.TH-3-95
hep-th/9504137
\end{flushright}
\newcommand{\ba}{\begin{array}}
\newcommand{\ea}{\end{array}}
\newcommand{\cR}{{\cal R}}
\newcommand{\lb}{\label}
\newcommand{\beqa}{\begin{eqnarray}}
\newcommand{\eeqa}{\end{eqnarray}}
\newcommand{\be}{\begin{equation}}
\newcommand{\ee}{\end{equation}}
\newcommand{\fr}{\frac}
\newcommand{\D}{\delta}
\newcommand{\sig}{\sigma}
\newcommand{\del}{\partial}
\newcommand{\wv}{\wedge}
\newcommand{\al}{\alpha}
\newcommand{\la}{\lambda}
\newcommand{\La}{\Lambda}
\newcommand{\ep}{\epsilon}
\newcommand{\pr}{\prime}
\newcommand{\ti}{\tilde}
\newcommand{\om}{\omega}
\newcommand{\OO}{\Omega_{BRST}}
\newcommand{\bi}{\bibitem}
\newcommand{\cA}{{\cal A}}
\newcommand{\cL}{{\cal L}}
\newcommand{\cD}{{\cal D}}
\newcommand{\cO}{{\cal O}}
\newcommand{\cN}{{\cal N}}
\newcommand{\Omo}{\Omega}

\begin{center}
{\large
THE BATALIN-VILKOVISKY METHOD OF
QUANTIZATION
MADE EASY }

\vspace{3cm}

{\normalsize  \"{O}mer F. DAYI}\footnote{E-mail address:
dayi@yunus.mam.tubitak.gov.tr}
\end{center}
{\small \it
TUBITAK - Marmara Research Center,
Research Institute for Basic Sciences,
Department of Physics,
P.O. Box 21,
41470 Gebze-TURKEY }

\vspace{2cm}
\begin{center}
{\bf  Abstract}
\end{center}

Odd time was introduced to formulate the Batalin-Vilkovisky
method of quantization of gauge theories in a systematic
manner. This approach is presented emphasizing  the
odd time canonical formalism beginning from an odd
time Lagrangian. To let the beginners have access to
the method essential notions of the gauge theories are
briefly discussed, and each step is illustrated with examples.
Moreover, the method of solving the master equation
in an easy way for a class of gauge theories is reviewed.
When this method is applicable some properties of the
solutions can easily be extracted as shown in the related
examples.

\vspace{3cm}

\pagebreak

\tableofcontents

\pagebreak

\section{Introduction}

Theoretical aspects of the elementary particles are discovered
in terms of gauge theories. The most common ones are
the Yang-Mills type theories whose quantization
is well understood. However,
especially after the construction of the covariant
string field theories (for a review see \cite{sft}), we learnt
that there are some other interesting gauge theories
which do not share the same properties with the Yang-Mills
theory: there may be open gauge algebras, and/or the gauge generators
may be linearly dependent (reducible gauge theory).
In the latter case it is sometimes possible to choose
a set of gauge transformations which behave like
the gauge transformations of Yang-Mills theory.
However, even if
this choice is possible usually it destroys a manifest
symmetry of the original system like
Lorentz invariance, which one prefers to keep.

The most efficient method of quantizing reducible gauge theories
whose gauge algebra is closed or open
is given by Batalin and Vilkovisky (BV)\cite{BVM}.
They offered a systematic way of finding the full
action which can be used in the related path integrals.
Unfortunately, this method appears
discouraging to the beginners
because the machinery used to formulate the method
is {\it ad hoc}: the reason of introducing antifields and
antibrackets is obscure.

The essential step in the application of the BV method
of quantization is to solve the (BV-) master equation.
However, obtaining the desired solution
is usually cumbersome. Moreover, the solutions are usually
very complicated, so that extracting some algebraic or
geometric properties of them is not easy.

There are some excellent reviews\cite{bvmr} and books\cite{bvmb}
on this subject, in which one can find some different applications
of the method as well as discussions
of some general aspects of it like the
structure equations resulting from the master equation.
However, the {\it ad hoc} definitions of the method are
kept, and there is no hint of solving
the master equation for complicated systems
in an easy way.

A solution to the former problem was given in terms
of the ``odd time" dynamics\cite{omot}, and
a general as well as an easy solution of the master equation
for a vast class of gauge theories is found\cite{omgs}, inspired
by the odd time formalism of the BV method.
However, a complete discussion of odd time Lagrangian and
the canonical formulation resulting from it was missing.
Moreover, neither the odd time dynamics nor the general solution
were presented in a complete and pedagogical manner.

The aim of this article is to present the BV method of quantization
without refering to any {\it ad hoc} definition
and to give an application of it
to a general class of  gauge theories,
in a way which renders easy
the access to the method and its applications.

In Section \ref{sfgt} first the basic concepts of
gauge theories are presented in terms of some examples.
The examples chosen are appropriate to illustrate
close and open gauge algebras and
linearly dependent (reducible) gauge generators. Then,
the reducibility conditions
suitable for applying the BV method of quantization are
given. Ghost and ghost of ghost
fields are introduced  in terms of path
integrals, and the Becchi-Roulet-Stora-Tyutin (BRST)
symmetry\cite{BRST} for an irreducible system is discussed.

We devote Section \ref{bvmq} to the BV method of quantization.
Odd time approach is presented
by discussing an odd time Lagrangian and the
related  canonical formalism in detail.
Then, a general solution of the master equation
which embraces a vast class of gauge theories
is discussed and applied
to the examples given in Section \ref{sfgt}.
Some properties of the proper solutions of the
examples are also presented.

\pagebreak

\section{Some Features of Gauge Theories}
\lb{sfgt}
\subsection{Generalities}
Here we briefly recall some properties of  gauge theories
illustrated by examples.

Let us deal with a theory given by
\be
\lb{ac1}
{\cal A}[\phi ]=\int d^dx {\cal L}(\phi , \del \phi / \del x ),
\ee
where the Grassmann parity of the fields are $\ep (\phi^i)=0$
(commuting) or $1$ (anticommuting) mod 2,
and $i=1, \cdots , n$.
It is supposed that the action possesses
at least one stationary point $\phi^i_0:$
\[
\fr{\D \cA}{\D \phi^i} |_{\phi^i_0}=0.
\]

For the sake of simplicity let us deal  with
bosonic $\phi^i$.
When the fields $\phi^i$ are transformed by some
infinitesimal local fields
$\al^a(x);$  $a=1,\cdots , m,$
\be
\lb{gt}
\D_\al \phi^i =R^i_a(\phi )\al^a (x),
\ee
if the action remains invariant
\be
\lb{tr}
\D_\al \cA = \int d^dx \fr{\D \cA}{\D \phi^i}
R^i_a \al^a = 0,
\ee
up to surface terms (or $\D_\al$exp$\cA =0)$,
the action (\ref{ac1}) defines a gauge theory. $R^i_a$ and $\al^a$
are gauge generators and gauge parameters.

We assume that all of the gauge transformations can be generated
by $R_a^i$, so that the commutator of two gauge transformations
can be written in terms of the generators
$R^i_a$, up to terms vanishing on
mass shell:
\be
\lb{algebra}
[ \D_\al ,\D_\beta ]\phi^i
\equiv \fr{\del R^i_{[a}}{\del \phi^j}R^j_{b]}\al^a \beta^b =
F^c_{ab}(\phi ) R_c^i\al^a \beta^b + \fr{\D \cA}{\D \phi^j}
K^{ji}_{ab} \al^a \beta^b .
\ee
Here, $[\ ]$ denotes antisymmetrization in the
indices which are within them.
If K vanishes identically, gauge transformations form
an algebra, and moreover if $F$ is independent of $\phi$
it is a Lie algebra. In the case where $K$ does not vanish,
gauge transformations still satisfy an algebra on mass shell,
hence it is called an open gauge algebra.

The generators
$R^i_a$, enumerated by $a$ can be
linearly independent or dependent. In the former case
the theory is named irreducible, and in the latter
case reducible theory. i.e. if there exists some (non-zero)
gauge parameters $\al_{(r)}^a$ satisfying
\[
R^i_a \al^a_{(r)}|_{\phi_{(0)}} = 0,
\]
the gauge theory is reducible.

Before discussing the conditions of reducibility in general,
which are adequate to use the BV method of quantization,
let us give some examples to illustrate
the cases discussed above.

\subsubsection{Examples}

\paragraph{Yang-Mills Theory:}

It is defined in terms of the second order action
\be
\lb{ym}
L_0= \fr{-1}{4}\int d^4x\   F^a_{\mu \nu}  F_a^{\mu \nu},
\ee
where in the differential form notation
$F=d \wedge A + A \wedge A$.
Gauge fields are in the adjoint representation of
the gauge group $SU(N),$ $A_\mu \equiv A_\mu^at_a,$
where
$t_a$ are the generators of the Lie algebra:
\[
{[t_a, t_b]}={f_{ab}}^ct_c.
\]
(\ref{ym}) is invariant under the gauge transformations
\be
\lb{ymg}
\D A_\mu =D_\mu \al\ ,
\ee
where $D=d + [A,\ ]$ is the covariant derivative. As one can easily
observe the theory is irreducible,
i.e. $D$ does not possess any non-trivial zero eigenvalue vector:
\be
\lb{t}
D^\mu \beta_k=0 \Longrightarrow \beta_k=0,
\ee
and the gauge transformations satisfy
the Lie algebra
\[
[\D_\al , \D_\rho ] A^a_\mu ={f_{bc}}^a D_\mu \al^b \rho^c.
\]

\paragraph{The Self-interacting Antisymmetric Tensor Field:}

The action\cite{ft} (we suppress $Tr$ which is over the
group indices, and  define ${\rm Tr}\  t_at_b =\D_{ab}$)
\be
\lb{si}
L_0 =-\int d^4x\   [B_{\mu \nu}(d\wv A+A\wv A)^{\mu \nu}
- \fr{1}{2} A_\mu A^\mu ],
\ee
is invariant under the transformations
\[
\D_\La B_{\mu \nu } =\ep_{\mu \nu \rho \sigma }D^\rho \La^\sigma ,\
\D_\La A_\mu =0 .
\]
Obviously, gauge algebra closes off shell
\be
{[\D_\La ,\D_\Sigma ]} (B_{\mu\nu},A_\mu )=0.
\ee

However, for $\La_\mu = D_\mu \al$, the gauge transformation
vanishes on shell $\D_\La B|_{F=0}=0$. In other terms the
gauge generators
\be
R_{\mu \nu}^\sigma =\ep_{\mu \nu \rho \sigma }D^\rho ,
\ee
possess
non-trivial zero eigenvalue vectors $D_\sigma$ on mass
shell:
\be
\lb{sil}
R_{\mu \nu}^\sigma D_\sigma |_{F=0} = 0.
\ee

Hence, this a reducible theory. Moreover, $D$ does not possess
non-trivial zero eigenvalue vector (\ref{t}).

\paragraph{Chern-Simons theory in $d=2p+1:$}

For $p=1,\ 2,\ 3\cdots ,$ it is given in terms of the action
\be
\lb{cs}
L_d= \fr{1}{2} \int_{M_d} \left( A \wedge dA
+\fr{2}{3} A\wedge A\wedge A \right) .
\ee

If the gauge field is  defined as
\be
A=\phi +\psi \equiv
\sum_{i=0}^{p-1} \phi_{2i+1}+\sum_{i=0}^p \psi_{2i},
\ee
where   $\phi_{2i+1}$ and $ \psi_{2i} $
are  Lie-algebra valued, respectively,
bosonic $2i+1$-form and fermionic $2i$-form,
the Chern-Simons action (\ref{cs}) yields
\be
\lb{sl}
L_d  =   \fr{1}{2} \int  _{M_d} \left( \phi \wedge d \phi
 +\fr{2}{3} \phi \wedge \phi \wedge \phi +
\psi \wedge D_\phi \psi \right),
\ee
where $D_\phi \equiv d +[ \phi ,\  ]$.
(\ref{sl}) is followed from the fact
that in the integral only the terms possessing odd differential
form degree
survive. This theory is introduced in
ref. \cite{mp}.

The action (\ref{sl}) is invariant under the gauge transformations
\be
\lb{rgt}
\D_\Sigma A= d\Sigma + [ A,\Sigma ] \equiv
\left(
\begin{array}{cc}
D_\phi & \psi \\
\psi   & D_\phi
\end{array}
\right)
\left(
\begin{array}{c}
\Lambda \\
\Xi
\end{array}
\right),
\ee
where the gauge parameter is
\be
\lb{116}
\Sigma = \La +\Xi \equiv \sum_{i=0}^{p-1} \La_{2i}
+\sum_{i=0}^{p-1} \Xi_{2i+1}.
\ee
$\La$ and $\Xi$ are bosonic and fermionic, respectively.
For some values of $\Sigma$ the gauge transformations
(\ref{rgt}) vanish on mass shell. Indeed, when the equations
of motion
\be
\lb{eqmm}
d\phi+[\phi ,\phi ] - \psi^2=0,\     D_\phi \psi =0,
\ee
are satisfied, the gauge generators generating (\ref{rgt})
are linearly dependent. Moreover the zero eigenvalue
vectors are also linearly dependent:
\be
\lb{zrmc}
Z_m Z_{m+1}=0,\ m=0, \cdots 2p-2,
\ee
where $Z_0$ is the gauge generator of (\ref{rgt}) and
\[
Z_{2m}=\left(
\ba{cc}
D_\phi &  \psi \\
\psi   &  D_\phi
\ea
\right) ,
Z_{2m+1}=\left(
\ba{cc}
D_\phi &  -\psi \\
-\psi   &  D_\phi
\ea
\right) .
\]

Observe the difference between the reducibility of this
theory and the previous one.

\paragraph{The Gauge Theory of Quadratic Lie Algebras:}

The algebra generated by $T_a$
\be
[T_a, T_b]={f_{ab}}^c T_c +V_{ab}^{cd}T_cT_d +k_{ab},
\ee
is  known as quadratic Lie algebra,
even if it is not a Lie algebra.
It is obtained by deforming the Lie algebra given by the
structure constants ${f_{ab}}^c.$ The constants
$f$, $V$, and $k$ possess the
symmetry properties
\be
\lb{sym}
{f_{ab}}^c=-{f_{ba}}^c,\   V_{ab}^{cd}=-V_{ba}^{cd},\
V_{ab}^{cd}=V_{ab}^{dc}, \  k_{ab}=-k_{ba}.
\ee
Moreover, they should be chosen to obey
\beqa
{f_{[ab}}^d{f_{c]d}}^e & =0, & \nonumber \\
{f_{[ab}}^dV_{c]d}^{ef} & +V_{[ab}^{df}{f_{c]d}}^e &+
V_{[ab}^{ed}{f_{c]d}}^f =0,  \nonumber \\
V_{[ab}^{de}V_{c]d}^{fg} & =0, &  \lb{ji} \\
{f_{[ab}}^d k_{c]d} & =0, &  \nonumber \\
V_{[ab}^{de}k_{c]d} & =0, &  \nonumber
\eeqa
because of the Jacobi identities.

Gauge theory of this algebra in 2--d space-time is given
by the Lagrange density\cite{II}
\be
\lb{l0}
{\cal L}=-\fr{1}{2}\ep^{\mu \nu}\{ \Phi_a(\del_\mu h_\nu^a
-\del_\nu h_\mu^a +{f_{bc}}^a h_\mu^bh_\nu^c +
V_{bc}^{ad} \Phi_dh_\mu^bh_\nu^c)+k_{ab}h^a_\mu h^b_\nu\},
\ee
which leads to the equations of motion
\beqa
\ep^{\mu \nu}(D_\nu \Phi_a +k_{ab}h_\nu^b) & = & 0,  \\
\ep^{\mu \nu}
(\del_\mu h_\nu^a
-\del_\nu h_\mu^a +{f_{bc}}^a h_\mu^bh_\nu^c +
2V_{bc}^{ad} \Phi_dh_\mu^bh_\nu^c) & = & 0 .
\eeqa
We used the definition
\[
D_\mu\Phi_a \equiv  \del_\mu \Phi_a +\Phi_cf^c_{ab}h_\mu^b +
\Phi_c\Phi_d V^{cd}_{ab}h_\mu^b.
\]

The action (\ref{l0}) is invariant under the gauge transformations
\beqa
\delta h_\mu^a & = & \del_\mu \la^a +{f_{bc}}^a h_\mu^b \la^c +2
V_{bc}^{ad} \Phi_dh_\mu^b\la^c, \lb{g1} \\
\delta \Phi_a & = & {f_{ba}}^c\Phi_c \la^b +
V_{ba}^{cd} \Phi_c \Phi_d \la^b +k_{ab}\la^b,  \lb{g2}
\eeqa
which satisfy
\beqa
[\D_\la ,\D_\eta ]h_\mu^a &  = &
\D_\kappa h_\mu^a -2\la^c \eta^d V^{ab}_{cd}
(D_\mu \Phi_b \lb{open1} +k_{be}h_\mu^e )\\
{[\D_\la ,\D_\eta ]}\Phi_a & = & \D_\kappa \Phi_a,
\eeqa
where
\[
\kappa^a \equiv ({f_{bc}}^a +
2V_{bc}^{ad} \Phi_d) \la^b \eta^c.
\]
Although,
the commutator (\ref{open1}) leads to an algebra only on mass
shell, the gauge generators of (\ref{g1})-(\ref{g2})
are linearly independent. Hence this theory is an example to an
irreducible gauge theory whose gauge generators satisfy an open algebra.

\subsection{Reducibility Conditions and Ghost Fields}
\lb{rcgf}

\subsubsection{Irreducible Gauge Theories}

When we deal with the partition function\footnote{We can equivalently
consider the  Green's functions generating functional
\[
Z[J]=\int [{\cal D}\phi^i] exp \int d^dx [ \cL +J\phi ],
\]
in terms of the gauge invariant sources $J_i$.}
of a gauge theory
\[
Z=\int [{\cal D}\phi^i] exp \int d^dx \cL ,
\]
the measure $[{\cal D}\phi^i],$ should take into consideration that
due to gauge invariance
some of the integrals
over fields are irrelevant and lead to infinities.
Eliminating these irrelevant degrees of freedom
usually causes destruction of some manifest symmetries
like covariance. Hence, one usually prefers to keep
all of the original fields, but put some gauge fixing conditions.
When the gauge generators $R_a$ are linearly independent (irreducible),
gauge fixing  can be achieved in terms of the conditions
\[
\chi_a (\phi) =0,
\]
whose Grassmann parity is denoted as
\[
\ep (\chi_a ) =\ep_a.
\]
Then the correct measure (or Haar measure) is
\[
[{\cal D}\phi^i] =\prod_{i,x} d\phi^i(x)\ \D (\chi_a(\phi))
{\rm det}^{(-)^{\ep_a}}
\left[
\fr{\del \chi_a}{\del \phi^i} \fr{\del (\D \phi^i)}{\del \al^b} \right],
\]
where $\al$ is the gauge parameter.
Let us define an effective action by putting the terms
which are in the measure into the exponent. To achive this
let us introduce the fields
\[
\la^a,\ \eta^a,\ \bar{\eta}^a;\  \ep (\la^a)=\ep_a\ ;\
\ep (\eta^a)=\ep (\bar{\eta}^a)=\ep_a+1\ .
\]
Now, in terms of them we  can write the
related path integral as
\be
Z =\int \prod_{i,x,a} d\phi^i(x) d\eta^a(x)d\bar{\eta}^a(x)
d\la^a(x) {\rm e^{-\cA_{eff}}},
\ee
where
\be
\lb{eff}
\cA_{eff} = \int d^dx \{\cL + \la^a\chi_a +\bar{\eta}^a
\left[
\fr{\del \chi_a}{\del \phi^i} \fr{\del(\D \phi^i)}{\del \al^b} \right]
\eta^b\}.
\ee
Obviously $\la^a$ are Lagrange multipliers and $\eta^a,\ \bar{\eta}^a$
are the so called ghost, antighost
fields. Observe that $\la_a$ possess the same
statistics but the ghosts $\eta^a$ and the antighosts $\bar{\eta}^a$
possess the opposite statistics of $\chi_a.$
Here, the ghosts $\eta^a$ are introduced as some auxiliary fields.
Hence, one should differ them from the original ones.
To this aim, introduce the ghost number ${\rm N_{ gh}},$
which is zero for the original
fields $\phi^i$ and the Lagrange multipliers $\la^a,$
but
\[
 {\rm N_{gh}} (\eta^a)= - {\rm {\rm N_{gh}}} (\bar{\eta}^a) =1.
\]

\subsubsection{Reducible  Gauge Theories}

As it is obvious from the above discussion, for a covariant
quantization some ghost fields are needed.
Batalin and Vilkovisky gave a way of performing this
for theories which satisfy some conditions\cite{BVM}:

\paragraph{First Stage Reducible Gauge Theories:}

Let us suppose that
\be
\lb{rz}
R^i_{a_0}{Z_1^{a_0}}_{a_1}|_{\phi_0}=0;
\ee
$ a_0=1,\cdots ,m_0;\ a_1=1,\cdots ,m_1<m_0,$
are satisfied by some non-trivial (nonzero) $Z_1$, but
$Z_{1a_1}$ are linearly independent.
Moreover, if
\beqa
{\rm rank}\ R^i_{a_0} & = & m_0-m_1<n;  \nonumber  \\
{\rm rank}\ {Z_1^{a_0}}_{a_1} & = & m_1; \nonumber \\
{\rm rank}\  \fr{\D^2  \cA }{\D \phi^i \D \phi^j}|_{\phi_0} & = &
n-(m_0-m_1),  \nonumber
\eeqa
are satisfied, the theory is a first stage reducible gauge theory.

Grassmann parities are denoted as
\[
\ep (R^i_{a_0}) =\ep_i+\ep_{a_0};\ \ep ({Z_1^{a_0}}_{a_1})
=\ep_{a_0} +\ep_{a_1}.
\]

Because of the linear dependence of $R_a$, not all of the original
gauge transformations are relevant.
To find an effective action for covariant quantization
we introduce
the zero stage ghost fields $\eta_0^{a_0}$, whose Grassmann
parity is $\ep_{a_0}+1$, and ghost number 1,
to write, similar to (\ref{eff}),
\be
\lb{eff0}
\cA^0_{eff} = \int d^dx \{ \cL + \la^{a_0}\chi_{a_0} +\bar{\eta}^{a_0}_0
\left[
\fr{\del \chi_{a_0}}{\del \phi^i} R^i_{b_0}
\right]
\eta^{b_0}_0 \}.
\ee
However, the transformations
\[
\D\eta_0^{a_0} ={Z^{a_0}_1}_{a_1}\al_1^{a_1},
\]
leave (\ref{eff0}) invariant. Now the ghost field $\eta_0$
behaves like a gauge field. So that, we introduce
some other ghosts (ghosts of ghosts), and antighosts
\[
\eta_1^{a_1}, \bar{\eta}_1^{a_1};
\ \ep (\eta_1^{a_1})=\ep (\bar{\eta}_1^{a_1})=\ep_{a_1}+1;
\  {\rm N_{gh}}(\eta_1^{a_1})=-{\rm N_{gh}}(\bar{\eta}_1^{a_1})=2,
\]
Lagrange multipliers and gauge fixing conditions, respectively,
\[
\la_1^{a_1},\ \chi^1_{a_1};\ \ep (\la_1^{a_1})=
\ep (\chi^1_{a_1})=\ep_{a_1};\
{\rm N_{gh}} (\la_1^{a_1})={\rm N_{gh}} (\chi^1_{a_1})
=0.
\]

We may choose the gauge fixing conditions $\chi^1_{a_1}$
depending  only on $\eta_0,$ so that the partition function
\be
Z^\prime =\int \prod_{i,x,a_0,a_1}
d\phi^i(x) d\eta_0^{a_0}(x)d\bar{\eta}_0^{a_0}(x) d\la_0^{a_0}(x)
d\eta_1^{a_1}(x)d\bar{\eta}_1^{a_1}(x) d\la_1^{a_1}(x)
{\rm e^{-\cA^\prime_{eff}}},
\ee
is written in terms of the effective action
\be
\lb{effp}
\cA^\prime_{eff} =\cA^0_{eff} +\int d^dx \{
\la_1^{a_1}\chi^1_{a_1} +\bar{\eta}^{a_1}_1
\left[
\fr{\del \chi^1_{a_1}}{\del \eta_0^{a_0}} {Z_1^{a_0}}_{b_1} \right]
\eta^{b_1}_1\}.
\ee

Therefore, the number of ghost fields depends on the level
of reducibility.

\paragraph{l$^{\rm th}$ Stage Reducible Gauge Theories:}

Suppose that in addition to (\ref{rz})
\[
{Z_r^{a_{r-1}}}_{a_r}
{Z_{r+1}^{a_r}}_{a_{r+1}}
|_{\phi_0}=0,
\]
$a_r=1, \cdots , m_r$,
are satisfied with non-trivial $Z_r$ for $r=1, \cdots ,l$, and
$Z_{l+1}=0$.
If they also satisfy
\beqa
{\rm rank}\ R^i_{a_0}  =  \beta_0 <n; &
\beta_0=\sum_{i=0}^l (-1)^im_i & \nonumber \\
{\rm rank}\ {Z_r^{a_{r-1}}}_{a_r}  = \beta_r; &  \beta_r=\sum_{i=r}^l
(-1)^im_i;\ & r=1, \cdots ,l \nonumber \\
{\rm rank}\ \fr{\D^2 \cA }{\D \phi^i \D \phi^j}|_{\phi_0}  =
n-\beta_0, & &   \nonumber
\eeqa
the theory is a l$^{\rm th}$ stage reducible gauge theory.

The Grassmann parities are denoted as
\[
\ep ({Z_r^{a_{r-1}}}_{a_r} ) =\ep_{a_{r-1}}+\ep_{a_r}.
\]

Similar to the previous case we need to introduce ghost and
ghost of ghost fields:
\[
\eta_r^{a_r};\ \ep (\eta_r^{a_r}) =\ep_{a_r};\  {\rm {\rm N_{gh}}}
(\eta_r^{a_r}) =r+1,\ r=0, \cdots , l-1.
\]
One can enlarge the set of fields by introducing
the related antighosts and Lagrange multipliers
in an obvious manner to discuss the effective action
similar to the previous cases.

To have an insight of the statistics of the ghost fields
let us assume that the original fields $\phi_i$, the
gauge generators $R$, and all of the $Z_r$ are bosonic.
Then, the zero stage ghosts $\eta_0$ are anticommuting, the
first stage ghosts $\eta_1$ are commuting and the rest is
continued by alternating Grassmann parity.

\subsection{Physical State Conditions in terms of the BRST Formalism }

Effective actions introduced contain some irrelevant
gauge fields, moreover some ghosts and auxiliary
fields which are not physical.
These fields should be isolated from the rest.
A way of performing this is the use of BRST symmetry. Before
explaining the method in general, we discuss this symmetry for
the simplest case.

Let us deal with a bosonic, irreducible gauge theory whose
gauge algebra closes off shell. Moreover, we suppose
that the gauge fixing functions $\chi_{a}$ are linear
in the fields $\phi_i,$
so that the effective action which can be used in path
integrals (\ref{eff}), becomes
\be
\lb{effir}
\cA_{eff} = \int d^dx [\cL + \la^a\chi_a +\bar{\eta}^a
\left[ \cO_{ai} (x) R_{b}^i \right]
\eta^b],
\ee
where $\cO_{ai} (x)$ are some operators.
This action is invariant under the  transformation
\be
\lb{0tr}
\D_Q\phi^i= R^i_a \eta^a,\
\D_Q \eta^a = -\fr{1}{2} F^a_{bc}(\phi ) \eta^b\eta^c,\
\D_Q \bar{\eta}^a = - \la^a,\
\D_Q \la^a = 0,
\ee
which is reminiscent of the gauge transformations
(\ref{gt})-(\ref{algebra}) and defined such that
\be
\D_Q^2(\phi ,\eta ,\bar{\eta},\la )=0.
\ee

$\D_Q$ is the well known BRST transformation\cite{BRST}.
As an example let us deal with
Yang-Mills theory.  In the covariant
gauge,
\[
\del^\mu A^a_\mu=0,
\]
the effective action reads
\be
\lb{ymeff}
\cA_{eff}^{YM}
=\int d^4x[ -\fr{1}{2}F_{\mu\nu}^2+\la\del^\mu A_\mu
+\bar{\eta}\del^\mu D_\mu\eta ].
\ee
Obviously, $\eta$ and $\bar{\eta}$ are fermionic fields.
One can observe that (\ref{ymeff}) possesses the symmetry
defined by
\be
\lb{trym}
\D_Q A_\mu=D_\mu\eta\ ;\ \D_Q\eta^a=-\fr{1}{2}f^a_{bc}\eta^b\eta^c\ ;\
\D_Q \bar{\eta}=-\la\ ;\ \D_Q \la=0.
\ee
Moreover, this transformation is nilpotent
\[
\D_Q^2(A,\la,\eta,\bar{\eta})=0.
\]

$\D_QA_\mu$ can be obtained from the gauge transformation (\ref{ymg})
by the replacement $\al \rightarrow \eta$.

A similar treatment of  reducible theories is also available.
Hence, let us deal with a gauge theory
given in terms of an effective Lagrangian ${\rm \cL_{eff} }(\Phi)$
and a nilpotent transformation $\D_Q$:
\be
\D_Q{\rm \cL_{eff}}(\Phi )=0,\ \D_Q^2\Phi_A=0,
\ee
where $\Phi_A$ denote the needed ghosts,
antighosts, Lagrange multipliers and the original fields.

By introducing the canonical momenta
\[
P_A=\fr{\del {\rm \cL_{eff}}}{\del (d\Phi_A/ dt)},
\]
one can write the BRST transformations in the phase
space:
\[
\D_Q \Phi_A =\Omega_A (\Phi , P),
\]
which leads to the Noether charge
\be
\lb{hc}
Q=\int d^{d-1}x[\Omega_A (\Phi ,P) P_A -K],
\ee
where $K$ is defined to satisfy
\be
\lb{K}
\fr{\del K}{\del P_A}-\fr{\del \Omega_A}{\del P_A} P_B=0.
\ee
By applying the usual canonical quantization procedure
one can find a nilpotent operator
\[
Q_{op}^2=0,
\]
resembling the charge (\ref{hc}), if there does not exist any
ordering anomaly.
Then, one defines the physical states as
\[
Q_{op}\psi_{phys.}=0,
\]
which are also defined to be on mass shell. In terms
of the perturbative analysis one can show that existence
of the charge $Q_{op}$ satisfying the above mentioned
properties is sufficient to eliminate the unphysical
degrees of freedom\cite{ko}.

Although, a charge which is related to $Q_{op}$ which
can be used to obtain the physical states is available
even before gauge fixing in terms of
the Batalin-Fradkin-Vilkovisky quantization scheme\cite{bfv},
it is out of the scope of this paper.

\pagebreak

\section{The Batalin-Vilkovisky Method of Quantization}
\lb{bvmq}

\subsection{Odd Time Formulation}

Inspired by supersymmetry, one can introduce a superpartner of time,
which we call ``odd time" $\tau $, satisfying
\[
\tau^2=0.
\]

Let us consider odd time dynamics in terms of the variable
$q_\mu (x, \tau )$, where $x$ indicates the usual time in
particle case, and all of the coordinates in field theory.
We use the same notation for the functions and the functionals.
Moreover, the integrals over $x$ are mostly suppressed.
In contrary to the usual mechanics, an implicit odd time dependence
does not make sense. One can always write
\[
q_\mu (x,\tau )= q_\mu (x,0) + \tau q_\mu^\prime (x,\tau ),
\]
where odd time derivative of $q_\mu $ is independent of $\tau $.
To emphasize this property we use the notation
\[
q_\mu^\prime (x,\tau )\equiv q_\mu^\prime (x).
\]
Hence, if we would like to describe the system in terms of
an odd time Lagrangian $L_o$, it will be in the following form
\be
\lb{otl}
L_o (q(x,0),q^\prime (x),\tau )=L_o (q(x,0),q^\prime (x),0)+\tau
L_o ^\prime (q(x,0), q^\prime (x), \tau ).
\ee
Obviously, $L_o ^\prime $ is independent of $\tau .$

Similar to
the usual case one can define ``odd time canonical momenta" as
\be
\lb{otcm}
p^\mu (x,\tau ) =
\fr{\del L_o (q(x),q^\prime (x),\tau ) }{\del q_\mu^\prime (x)}.
\ee
$L_o $ is supposed to be bosonic, so that
$p^\mu $ possesses the opposite
statistics of $q_\mu .$ Hence, we deal with
a supermanifold
possessing an equal number of fermionic and bosonic coordinates.
On such a manifold
there exist
an even as well as an odd canonical two form.
We would like to deal with the latter one\cite{ocf}.

Only the first derivative of a function with respect to odd time
can be non-vanishing.
Thus, for a canonical formalism it is sufficient to
discuss only first order Lagrangians.
In general
one can deal with the two sets of variables
\[q_\mu \equiv
(a_i(x,\tau ),b_i(x,\tau ))=(a_i(x,0 )+ a^\prime (x)\tau ,
b_i(x,0)+\tau b^\prime_i(x)).
\]
We choose odd time Lagrangian to be
\be
\lb{L_o}
L_o=a_i(x,0)b_i^\prime (x)+ a_i^\prime (x) b_i(x,0)
+ a_i^\prime (x)\tau b_i^\prime (x) - S(a_i(x,0),b_i(x,0)).
\ee
Grassmann parity of the variables should be
$\ep (a_i)=\ep (b_i)+1.$
Odd time canonical momenta are defined as
\beqa
p_a^i & = & \fr{\del_l L_o}{\del a^\prime_i} = b_i (x, \tau ),\lb{opa} \\
p_b^i & = & \fr{\del_r L_o}{\del b^\prime_i} = a_i (x, \tau ), \lb{opb}
\eeqa
where right and left derivatives are related
\[
\fr{\del_rf(z)}{\del z}=(-1)^{\epsilon (z)[\epsilon (f)+1]}
\fr{\del_lf(z)}{\del z}.
\]
$L_o $ and right and left derivatives in the definitions of
odd canonical momenta (\ref{opa})-(\ref{opb}),
are chosen to avoid $(-1)$ factors.

We define the related odd Hamiltonian as
\beqa
H_o & \equiv & a_i^\prime p^i_a +p_b^ib^\prime_i
-a_i^\prime \tau b_i^\prime - L_o, \\
 & = &  S(a_i(x,0),b_i(x,0)).
\eeqa
In terms of this definition odd time independence of $H_o$
is guaranteed:
\be
\lb{hp0}
\fr{\del H_o}{\del \tau }=0.
\ee

Now, one can define an ``odd Poisson bracket"(antibracket)
in terms
of $ a,b$ and their canonical momenta. Because of the
constraints (\ref{opa}), (\ref{opb}) one can eliminate
$p_a,\ p_b$ such that the basic odd Poisson brackets are
\be
(a_i,b_j) =\delta_{ij} .
\ee
i.e.
Observables $f,\ g$ are functions of $a_i,\ b_i$ and their
odd Poisson bracket is
\be
(f,g)=\fr{\del_r f}{\del b_i}\fr{\del_l g}{\del a_i} -
\fr{\del_r f}{\del a_i}\fr{\del_l g}{\del b_i}.
\ee
The odd Poisson bracket (antibracket) has the following properties
\be
\label{eq:A}
\epsilon [(f,g)] = \epsilon (f) + \epsilon (g)+1 ,
\ee
\be
\label{eq:B}
(g,f)=-(-1)^{[\epsilon (f) +1][\epsilon (g) +1]}(f,g),
\ee
\beqa
(-1)^{[\epsilon (f)+1][\epsilon (g)+1]}(g,(l,f)) & + &
(-1)^{[\epsilon (g)+1][\epsilon (l)+1]}(l,(f,g)) \nonumber \\
 & + &
(-1)^{[\epsilon (l)+1][\epsilon (f)+1]}(f,(g,l))=0. \label{eq:C}
\eeqa

We should clarify the meaning of the derivative with respect to
$q_\mu (x,\tau ).$ It is demanded to satisfy
\be
\lb{cqdq}
\fr{\del}{\del q_\mu }q_\nu -q_\nu \fr{\del}{\del q_\mu }=
\delta_\nu^\mu .
\ee
The choice
\be
\lb{ch}
\fr{\del}{\del q_\mu (x,\tau ) } =
\fr{\del}{\del q_\mu (x,0 ) } +\tau
\fr{\del}{\del q^\prime_\mu (x ) } ,
\ee
can be seen to satisfy (\ref{cqdq})
in the space of functions which are polynomials in $q_\mu .$

Similar to
the usual case let the odd  time evolution of an observable
$f$ is generated by the odd time Hamiltonian $S$ in terms of
the odd Poisson bracket
\be
\lb{hcch}
f^\prime (a,b)\equiv (S,f)=
\fr{\del_r S}{\del b_i(x,0)}
\fr{\del_l f}{\del a_i (x,\tau )}-
\fr{\del_r S}{\del a_i(x,0)}
\fr{\del_l f}{\del b_i (x,\tau )},
\ee
where we used (\ref{ch}).

Equations of motion of the canonical variables are
\beqa
a_i^\prime & =(a_i,S) &=-\fr{\del_l S(a(x,0),b(x,0))}{\del
b_i (x,0 )},  \lb{br1} \\
b_i^\prime & =(b_i,S)= &\fr{\del_l S(a(x,0),b(x,0))}
{\del a_i (x,0 )}. \lb{br2}
\eeqa
Observe that these agree with
odd time equations of motion
resulting from the odd time Lagrangian (\ref{L_o}),
if they are defined as
\beqa
\fr{\del_l L_o}{\del a_i (x,\tau ) }
- \tau \fr{\del_l L_o}{\del a^\prime_i (x) } = 0, \lb{eqm1} \\
\fr{\del_r L_o}{\del b_i (x,\tau ) }
-\tau \fr{\del_r L_o}{\del a^\prime_i (x) } = 0. \lb{eqm2}
\eeqa
In (\ref{eqm2}) agreement is up to a sign factor if $b$ is bosonic.

Moreover, $S(a(x,0),b(x,0))$ should be invariant under the odd time
evolution (\ref{hcch}):
\be
\lb{ss0}
(S,S)=
\fr{\del_r S}{\del a_i(x,0)}
\fr{\del_l S}{\del b_i (x,0 )}-
\fr{\del_r S}{\del b_i(x,0)}
\fr{\del_l S}{\del a_i (x,0 )} =0,
\ee
which is known as (BV) master equation.
By making use of (\ref{eq:C})
and (\ref{ss0}) one can show that
the second derivative of an observable $f$ with respect
to the odd time is vanishing
\[
\fr{\del^2f}{\del \tau^2}=(S,(S,f))=0.
\]

In contrary to the usual case,
invariance of Hamiltonian under the odd time evolution
(\ref{ss0}) is not trivially satisfied.

To attribute a physical content to the odd time dynamics
one should specify
the fields $a,b$ and meaning of the odd time
evolution (\ref{hcch}).
Here we use this formalism to formulate the BV method of
quantization of gauge theories.

Let us deal with a gauge theory given in terms of an action
$\cA (\phi )$ invariant under the gauge transformations (\ref{gt}).
Then, we identify the derivative with respect to
the  odd time  with the BRST transformation (or charge):
\be
\lb{eod}
\fr{\del}{\del \tau} \equiv \delta_{BRST}.
\ee
Now, analyse the reducibility of the gauge generators
$R_a^i$ and introduce
the needed ghost fields possessing positive ghost number,
as outlined in Section \ref{rcgf}.
Hence we assign
\be
N_{{\rm gh}}(\fr{\del}{\del \tau} )=1.
\ee

If the odd time Lagrangian
(or Hamiltonian) (\ref{otl}) is physical, it has to satisfy
\be
N_{ {\rm gh}} (L_o )=0.
\ee
Thus, to write an odd time Lagrangian which depends on  ghost fields
one should introduce some other fields possessing negative
ghost number. Now, if $q_i$ denote  the original  and
the ghost fields, and $p_i$ the odd time canonical momenta
defined as (\ref{otcm}), they should satisfy
\be
N_{ {\rm gh}}(p_i)= N_{ {\rm gh}}(L_o)-(N_{ {\rm gh}}(q_i) +1) ,
\ee
which leads to
\be
\lb{-1}
N_{ {\rm gh}}(p_i)+N_{ {\rm gh}}(q_i) =-1.
\ee
Therefore, the number of
positive and negative ghost number components
of the fields $a,\ b$ used to write $L_o$  should be the same.
To simplify the notation as well as to connect it to the usual one,
let us rename the odd time independent components of $a$ and $b:$
\[
a_i(x,0)\equiv \Phi_i (x);\  b_i(x,o) \equiv \Phi_i^\star (x),
\]
such that
\[
N_{ {\rm gh}}(\Phi_i ) \geq 0,\  N_{ {\rm gh}}(\Phi_i^\star ) < 0.
\]
Moreover, they should satisfy (\ref{-1}), namely
\be
\lb{cgn}
N_{ {\rm gh}}(\Phi_i)+N_{ {\rm gh}}(\Phi^\star_i) =-1.
\ee
The master equation (\ref{ss0}) is now
\be
\lb{ss1}
(S,S)=2 \fr{\del_r S}{\del \Phi_i} \fr{\del_l S}{\del \Phi^\star_i} =0,
\ee
and the BRST transformations (\ref{br1})-(\ref{br2}) are given by
\be
\lb{br}
\D_{BRST} \Phi_i = \fr{\del_l S}{\del \Phi^\star_i},\
\D_{BRST} \Phi_i^\star = -\fr{\del_r S}{\del \Phi_i} .
\ee

$\Phi^\star_i$ are known as antifields.

Till now we specified the field content of the formalism.
The second link to the usual notions of field theory is
to demand that ``the classical limit'' of the odd time
Hamiltonian is
\be
S(\Phi ,\Phi^\star )|_{\Phi^\star =0}= \cA [  \phi ].
\ee

Let the total number of the phase space variables $(\Phi , \Phi^\star )$
is denoted by $2N.$
$N$ of them are ``unphysical''(from the odd time formulation
point of view) because they are introduced as
odd canonical conjugates.
On the other hand, by taking the derivative of
(\ref{ss1}) one obtains
\be
\left( \fr{\del_r S}{\del \Phi_i}\ ,\ \fr{\del_r S}{\del \Phi_i^\star }
\right) {\cR}_{ij}=0,
\ee
where
\beqa
{\cR}_{ij} & =  &
\left(
\begin{array}{cc}
\fr{\del_l \del_r S}{\del \Phi_i^\star \del\Phi_j} &
\fr{-\del_l \del_r S}{\del \Phi_i^\star \del\Phi_j^\star} \\
\fr{\del_l \del_r S}{\del \Phi_i \del\Phi_j} &
\fr{-\del_l \del_r S}{\del \Phi_i \del\Phi_j^\star}
\end{array}
\right) .        \lb{rij}
\eeqa
Thus, $S$ is invariant under the gauge transformations generated by $\cR .$
If $\cR$ satisfies
\be
\lb{rN}
{\rm rank}\ \cR_{ij} =N,
\ee
gauge invariance permits us to eliminate the undesired variables.
Moreover $\cR$ satisfies
\[
\cR_{ij}\cR_{jk}=0,
\]
so that, $N$ is its maximal rank, which ensures that
all of the gauge invariances are taken into account.

Solution of the master equation (\ref{ss1}) satisfying (\ref{rN})
is called proper.

By expressing the ``unphysical'' variables $\Phi^\star$
in terms of the ``physical'' ones $\Phi$
one can fix the gauge invariance. However, the conditions on their
ghost numbers (\ref{cgn}) do not allow this.
Therefore, to achive
gauge fixing in this way one should enlarge the space of the
original and ghost fields by introducing
\[
\Sigma_z,\ \Lambda_z;\   N_{ {\rm gh}}(\Sigma_z) =N_{ {\rm gh}}(\Lambda_z)-1
=-N_{ {\rm gh}}(\Phi_z) ,
\]
where $\Phi_z$ indicate the fileds $\Phi_i$ except the original
gauge fields.
Of course, for not altering the number of the physical variables
one should define a new solution of master equation as
\be
\lb{exts}
S_e (\Phi^A ,\Phi^\star_A) =S(\Phi_i ,\Phi_i^\star)
+\Lambda_z\Sigma^\star_z,
\ee
where
\[
\Phi^A \equiv (\Phi_i , \Sigma_z , \Lambda_z ) .
\]
Now, gauge fixing can be obtained as
\be
\Phi_A^\star =\fr{\del \Psi (\Phi^A )}{\del \Phi^A}.
\ee
Obviously, $\Psi (\Phi )$ should be fermionic and moreover,
it should possess
\[
N_{ {\rm gh}}(\Psi )=-1.
\]
Because of these properties $\Psi$ is called ``gauge fixing fermion".

The gauge fixed action
\be
S_e (\Phi^A ,\del\Psi / \del \Phi^A ) =S(\Phi_i ,\del \Psi / \del\Phi_i)
+\Lambda_z \fr{\del \Psi  }{\del \Sigma_z},
\ee
can be used in the related path integral
(partition function)
\be
Z=\int \cD \Phi^A {\rm exp}
\{ S_e (\Phi^A ,\del\Psi / \del \Phi^A ) \} ,
\ee
or in the Green's function generating functional.

\subsubsection{Solution of the Master Equation for Yang-Mills Theory}

Before proceeding with the general formalism, to illustrate
the method let us apply it
to  Yang-Mills theory.

This theory is described in terms of the action
(\ref{ym}), which possesses the gauge symmetry given in (\ref{ymg}).
Because of being an irreducible gauge theory, we introduce only
the zero stage ghosts $\eta^a,$ which are anticommuting, and
possessing ghost number $1.$
The minimal set of fields is
\[
\Phi_i =(A_\mu^a,\ \eta^a).
\]
The odd canonical conjugates (antifields) are
\[
\Phi_i^* =(A_\mu^{a*},\ \eta^*_a);\  \ep (A^*)=1,\ \ep (\eta^*)=0;\
N_{ {\rm gh}} (A^\star )=-1, N_{ {\rm gh}} (\eta^\star )=-2.
\]
The proper solution of the master equation
(\ref{ss1}) can easily be obtained  as
\be
\lb{ymps}
S^{YM}(\Phi_i,\Phi^\star_i)=\int d^4x[ -\fr{1}{4}F_{\mu\nu}^{a}
F^{\mu\nu}_{a}
+A^{a*}_\mu  (D^\mu\eta )_a -\fr{1}{2} \eta^*_af^a_{bc}\eta^b \eta^c].
\ee
For  gauge fixing we enlarge the set of fields by
\beqa
\bar{\eta}^a,\bar{\eta}_a^*;\  \la^a,\la^*_a; & &\nonumber \\
\ep(\bar{\eta}^*)=\ep(\bar{\eta}) +1=0,\  &
\ep (\la )=\ep (\la^*) +1 =0; & \nonumber \\
N_{ {\rm gh}} (\bar{\eta} )=N_{ {\rm gh}} (\la^\star )=-1, \ &
N_{ {\rm gh}} (\bar{\eta}^\star )=N_{ {\rm gh}} (\la )=0. & \nonumber
\eeqa
The extended proper solution of the master equation (\ref{exts}) is
\[
S^{YM}_e(\Phi^A,\Phi^\star_A) =S^{YM}(\Phi_i,\Phi^\star_i)
-\int d^4x\ \bar{\eta}^*_a \la^a .
\]
{}From this action one can read the BRST transformations
by using the definition
(\ref{br}), and observe that they are the same
with (\ref{trym}).
We choose the gauge fixing fermion as
\[
\Psi =-\bar{\eta}^a \del_\mu A^\mu_a ,
\]
so that the gauge fixed action
\[
S^{YM}_e(\Phi^A,\del \Psi / \del \Phi^A)
=\int d^4x[ -\fr{1}{4}F_{\mu\nu}^2+\la\del^\mu A_\mu
+\bar{\eta}\del^\mu D_\mu\eta ],
\]
coincides with (\ref{ymeff}).

\subsection{Quantum Master Equation}

The extended solution $S_e$ found by odd time approach is a classical
action. The action after taking into consideration
quantum corrections will be
\[
W(\Phi , \Phi^*)=S_e (\Phi , \Phi^*) +\sum_{n=1}^\infty
\hbar^nW_n(\Phi , \Phi^*).
\]
The gauge fixed action
\[
W(\Phi , \fr{\del \Psi (\Phi )}{\del \Phi} ) \equiv
W(\Phi , \Phi^*)|_\Sigma,
\]
can be used in the partition function
\be
\lb{pf}
Z=\int \prod_{x,A}d\Phi^A
{\rm exp}[-i\hbar W(\Phi ,\Phi^\star )|_\Sigma ] .
\ee
Let the BRST transformation is still given by
\be
\lb{pb}
\D_{BRST} \Phi^A \equiv
(W, \Phi^A)|_\Sigma  =
\fr{ \del W}{ \del \Phi^*_A}|_\Sigma .
\ee
Hence the partition function transforms as
\[
\D_{BRST} Z = \int \prod_{x,A}d\Phi^A
\left[ \fr{ \del_r}{ \del \Phi^A}
(\fr{ \del_l W}{ \del \Phi^*_A}|_\Sigma )
+\fr{i}{\hbar}\fr{ \del_r W|_\Sigma }{ \del \Phi^A}
\fr{ \del_l W}{ \del \Phi^*_A}|_\Sigma
\right]
{\rm exp}[-i\hbar W|_\Sigma ] .
\]
Here the former term in the parenthesis is due to the
change in the measure. If one demands
invariance of the partition function under the BRST
transformation (\ref{pb}),
\be
\fr{-i}{2\hbar}  (W,W)+\Delta W +O_1+O_2 =0,
\ee
should be satisfied. Here we used the operator
\[
\Delta =\fr{ \del_r  \del_l}{ \del\Phi^A  \del \Phi^*_A},
\]
and the terms $O_1$, and $O_2$ are
\beqa
O_1 &  = &  \fr{ \del_r W}{ \del \Phi_B^*}
\fr{ \del^2_r\Psi}{ \del \Phi^A \del \Phi^B }
\fr{ \del_r W}{ \del \Phi_A^*},\nonumber \\
O_2  &  =  &  \fr{ \del_r \del_lW}{ \del \Phi^*_B  \del\Phi^*_A}
\fr{ \del^2_r \Psi }{ \del \Phi^A  \del \Phi^B}. \nonumber
\eeqa
By using the symmetry properties one can show that
\[
O_1=-O_1=0;\   O_2=-O_2=0.
\]
Hence if $W$ satisfies the equation
\be
\lb{qme}
\fr{1}{2} (W,W)+i\hbar \Delta W =0,
\ee
the partition function (\ref{pf}) is invariant under the BRST
transformation (\ref{pb}). (\ref{qme}) is known as the quantum master
equation. Now by expanding it in powers of $\hbar$ one obtains
\beqa
(S,S)  = & 0,   & \lb{qm1} \\
(W_1,S)  = & i\Delta S, &    \\
(W_n,S)  = & i\Delta W_{n-1} &
-\fr{1}{2} \sum_{m=1}^{n-1} (W_m,W_{n-m}).  \lb{qm2}
\eeqa
The  terms added after enlarging the minimal set of fields are
such that they identically satisfy the master equation and
moreover, they do not give any contribution to the quantum corrections
of the action. Hence,
in (\ref{qm1})-(\ref{qm2}) we can drop them and
consider only the minimal
set of the fields $\Phi_i$

\subsection{A General Solution of the Master Equation}
\lb{sgs}

A proper solution of the master equation can be
found by writing  $S$
as a polynomial in antifields. This does not cause
any difficulty for simple systems like Yang-Mills theory
or antisymmetric tensor field. However,
usually  applying this procedure
is complicated for the systems whose
level of reducibility is high and/or possess an open gauge algebra.
Moreover, a geometrical or algebraic
interpretation of the solutions is obscure.
Here, we present an easy solution which can be applied
to a vast class of gauge theories. Moreover, it is
suitable to extract some geometric or algebraic
properties of the quantized theory.

One may treat the exterior derivative $d$  and
the derivative with respect to the odd time on the same
footing by introducing
\be
\lb{dtil}
\ti{d} \equiv d + \del / \del \tau .
\ee
In terms of the identification (\ref{eod}) $\ti{d}$ can equivalently
be written as
\be
\lb{dBr}
\ti{d} \equiv d + \D_{BRST} ,
\ee
which is defined to satisfy\cite{BTM},\cite{nedall}
\[
\ti{d}^2= d \D_{BRST} +\D_{BRST} d=0.
\]
Recall that the exterior derivative $d$ and
the BRST transformation
$\D_{BRST}$ increase, respectively,
differential form degree and ghost number by one.

Consider the minimal set of  fields and antifields needed
in the BV method of quantization i.e. the original fields,
ghosts, ghosts of ghosts, and their antifields.
Their main distinguishing
parameters are: $i )$ghost number, $ i i )$ behavior
under the Lorentz transformations i.e.
differential form degree.
These two different properties can be unified in terms
of the generalized derivative (\ref{dBr}). i.e. by considering
the total degree
\be
\lb{tde}
\cN \equiv N_d +N_{\rm gh},
\ee
where $N_d$ denotes differential form degree.

A general solution will be given for the gauge systems
whose  Lagrangian (action) can be put into the first order form
\be
\lb{ola}
L(A,B)= BdA + V(A,B).
\ee
Gauge transformations can be written as
\be
\lb{caz}
\D^{(0)} (A,B) = R^{(0)} (A,B)\La ,
\ee
by suppressing the indices.

The minimal set of fields
can be figured out
analysing  reducibility  of the  gauge transformations
(\ref{caz}), as discussed in Section \ref{rcgf}.
They can be collected in groups as $\ti{A},$ and $\ti{B}$ satisfying
\be
\lb{cto}
\cN (\ti{A}) = \cN (A)\ ;\ \cN (\ti{B}) = \cN (B).
\ee
If differential form degrees of the original fields are
various, the above mentioned generalization should be done for each
degree. Then,
substitute the original fields
$A,B$ with the generalized ones
$\ti{A},  \ti{B}  $ in the Lagrangian (\ref{ola}):
\be
\lb{mass}
S \equiv L(\ti{A}, \ti{B})= \ti{B}d\ti{A} +V(\ti{A},\ti{B} ).
\ee
In (\ref{mass}) multiplication is defined such that
$S$ is a scalar possessing zero ghost number:
\[
\cN (S)= N_d(S)= N_{\rm gh} (S)=0,
\]
(\ref{mass}) is invariant under the transformations
\be
\lb{nonso}
\D_{\ti{\La}} (\ti{A},\ti{B}) = \ti{R} \ti{\La},
\ee
where the generators are
\be
\lb{ggg}
\ti{R}\equiv R^{(0)}(\ti{A} , \ti{B} ),
\ee
and $\ti{\La}$ is the appropriate generalization of the
gauge parameter $\La :$
\[
\cN (\ti{\La})=\cN (\La ).
\]
$S$ given by (\ref{mass}) is
the solution of the master equation if (\ref{nonso}) can be
written as
\beqa
\left(
\begin{array}{c}
\D_{\ti{\La}} \ti{A}_i \\
\D_{\ti{\La}} \ti{B}_i
\end{array}
\right)   & = &
\left(
\begin{array}{cc}
-\fr{\del_l\del_rS}{\del \ti{B}_i \del \ti{A}_j} &
- \fr{\del_l\del_rS}{\del \ti{B}_i \del \ti{B}_j} \\
\fr{\del_l\del_rS}{\del \ti{A}_i \del \ti{A}_j} &
\fr{\del_l\del_rS}{\del \ti{A}_i \del \ti{B}_j}
\end{array}
\right)
\left(
\begin{array}{c}
\ti{\La}_1^j  \\
\ti{\La}_2^j
\end{array}
\right),
\lb{ggi}
\eeqa
with $(\ti{\La}_1\neq 0,$ $\ti{\La}_2 \neq 0)$
or $(\ti{\La}_1\neq 0,$ $\ti{\La}_2 = 0)$
or $(\ti{\La}_1= 0,$ $\ti{\La}_2 \neq 0)$.

Variation of
$S$ under (\ref{ggi}) can be shown to yield
\be
\D S = \fr{\del_r (S,S)}{\del \ti{A}_j} \ti{\La}_1^j
+\fr{\del_r (S,S)}{\del \ti{B}_j} \ti{\La}_2^j .
\ee

Obviously, when
$\ti{\La}_1\neq 0$, and $\ti{\La}_2 \neq 0$,
$S$
satisfies the master equation, and $\ti{R}$
coincides with (\ref{rij})\footnote{$(S,S)={\rm const.} \neq 0$
would lead to the non-consistency of the equations of motion.}.
The same conclusion can be derived when one of the
parameters $\ti{\La}_{1}$ or $\ti{\La}_{2}$ is vanishing.
Let $\ti{\La}_1 \neq 0$, $\ti{\La}_2=0,$ so that,
$(S,S)$ is independent of $\ti{A}$.
$(S,S)$ possesses $N_d=0,\ N_{\rm gh} =1,$
indicated as $(0,1).$
However, usually it is not possible to
construct a function possessing $(0,1)$ degree
only in terms of $\ti{B}$. Hence, we can conclude that
$(S,S)$ vanishes.
The other case, $\ti{\La}_1 = 0$, $\ti{\La}_2\neq 0$,
can be examined similarly.

We choose the signs of the field contents of
$\ti{A}$ and $\ti{B}$ as
\[
\ti{A}_i=(\Phi_k ,\Phi_l^*),\      \ti{B}_i=(-\Phi_k^*, \Phi_l),
\]
so that, the transformations
\be
\lb{bt}
\D \ti{A}_i =\fr{\del_r S}{\del \ti{B}_i},\
\D \ti{B}_i =-\fr{\del_r S}{\del \ti{A}_i},
\ee
define the BRST transformations in accordance with the BV formalism
(\ref{br}). Obviously, in (\ref{bt}) the right hand side is defined
to have one more ghost number, but the same $N_d$
of the field appearing on the left hand side.

By construction $S(\ti{A},\ti{B} )$ possesses the correct classical
limit:
\[
S|_{\Phi^\star =0} =L(A,B).
\]
Moreover, in $\ti{A}$ and $\ti{B}$
all the fields of the minimal sector are included, and
because of the form of $S$, (\ref{mass}),
\[
{\rm rank}\  \left| \fr{\del^2 S}{\del (\ti{A},\ti{B}) \del
(\ti{A},\ti{B})} \right| = N,
\]
where $N$ is the number of the components of
$\ti{A} $ or $\ti{B} $. This is the condition given in
(\ref{rN}). Hence, we conclude that
under the above mentioned conditions $S=L(\ti{A},\ti{B})$
is the proper solution of the master equation.

\subsection{Examples to the General Solution}

\subsubsection{Yang-Mills Theory}
\lb{ex1}

The first order action
\be
\lb{eym}
L=\fr{-1}{2}\int d^4x\   ( B _{\mu \nu}   F^{\mu \nu}
-\fr{1}{2} B_{\mu \nu} B^{\mu \nu} ),
\ee
is equivalent to  (\ref{ym}) on mass shell, and it
is invariant under the
infinitesimal gauge transformations
\[
\D A_\mu =D_\mu \La\ ,\  \D B_{\mu \nu}= [ B_{\mu \nu}, \La ].
\]
The theory is irreducible, so
that for the covariant quantization we
need to introduce (in the minimal sector) the ghost field
$\eta_{(0,1)}$, and the antifields $A^\star_{(3,-1)},$
$\eta^\star_{(4,-2)},$ and $B^\star_{(2,-1)}.$
The first number in parenthesis is the differential form
degree and the second is the ghost number. Here the star indicates
the antifields as well as the Hodge-map.

By using (\ref{cto})
we write the generalized fields as
\[
\begin{array}{lcl}
\ti{A} & =  &  A_{(1,0)}+\eta_{(0,1)} + B_{(2,-1)}^\star , \\
\ti{B} & = & - A^\star_{(3,-1)}- \eta^\star_{(4,-2)}+B_{(2,0)}.
\end{array}
\]

In terms of the substitution
\[
A  \rightarrow \ti{A}, \ B \rightarrow \ti{B},
\]
in (\ref{eym}) one obtains
\be
\lb{obt}
S=\fr{-1}{2}\int d^4x\   [\ti{B} (d\ti{A}
+\ti{A}\ti{A}) -\fr{1}{2} \ti{B}\ti{B} ].
\ee

By using the property of the multiplication
that the scalar product is different
from zero only when its ghost number vanishes, we
get
\be
\lb{syme}
S=-\int d^4x\ (\fr{1}{2}B_{\mu \nu}F^{\mu \nu} -
B^{\mu\nu} [\eta , B_{\mu\nu}^\star ]
+A_\mu^\star D^\mu \eta +
\fr{1}{2}\eta^\star [\eta ,\eta ]
-\fr{1}{4} B_{\mu\nu} B^{\mu\nu}).
\ee
We may perform a partial gauge fixing $B^\star =0$, and then
use the equations of motion with respect to $B_{\mu\nu}$ to obtain
\[
S\rightarrow S^{YM}= \int d^4x\   (-\fr{1}{4}F_{\mu \nu}  F^{\mu \nu}
+A^\star_\mu    D^\mu \eta
-\fr{1}{2}\eta^\star [\eta ,\eta] ),
\]
which is the minimal solution of the master equation for
Yang-Mills theory (\ref{ymps}).
One can also observe that
(\ref{syme}) is the desired solution
by denoting that indeed,
the transformations (\ref{bt}) are the BRST transformations:
using (\ref{obt}) in (\ref{bt}) leads to
\be
\lb{bz}
\delta \ti{A}  = \ti{F}  -\ti{B}  ,\  \delta \ti{B}  = -\ti{D} \ti{B}  ,
\ee
where $\ti{D} = d+[\ti{A},] $ and  $\ti{F}$ is the related curvature.
Formally we have
\[
\begin{array}{lll}
\delta^2 \ti{A}   & = &  \ti{D} \cdot  (\ti{F} -\ti{B} )+\ti{D}
\ti{B} =0 , \\
\delta^2\ti{B}  & = & -\ti{D}  \cdot  \ti{D} \ti{B}
+ (\ti{F} -\ti{B} )\cdot  \ti{B}=0 ,
\end{array}
\]
due to the Bianchi identities $\ti{D} \cdot  \ti{F} =0$,
the definition of the curvature $\ti{F} =\ti{D} \cdot  \ti{D} $, and
$\ti{B} \cdot  \ti{B}
=\ti{B} _i\ti{B} _j - (-1)^{\ep (\ti{B} _i) \ep (\ti{B} _j)}
\ti{B} _j \ti{B} _i =0. $

In the gauge $B^\star =0$ use of the equations of motion
$B_{\mu \nu}=F_{\mu \nu}$ in (\ref{bz}) yields
\[
(\D + d)(A+\eta ) +[(A+\eta ),(A+\eta )  ] =F
\]
which is the Maurer-Cartan horizontality condition\cite{BTM}.

\subsubsection{The Self-interacting Antisymmetric Tensor Field}

As we have seen, (\ref{si})-(\ref{sil}), this system
is a first stage
reducible theory. Hence we need to introduce the
noncommuting ghost and commuting ghost of ghost fields
\[
C_0^\mu , \ C_1 ;\  {\rm N_{gh}}(C_0^\mu )=1 , \ {\rm N_{gh}}(C_1)=2.
\]
After introducing the related antifields, the generalized fields can be
written as
\[
\begin{array}{ll}
\ti{A} = &  A_{(1,0)} +B^\star_{(2,-1)} +C_{0(3,-2)}^\star
+C_{1(4,-3)}^\star ,\\
\ti{B} = & -A^\star_{(3,-1)}+ B_{(2,0)}  +C_{0(1,1)} +C_{1(0,2)}.
\end{array}
\]

By following the general procedure we find
\be
\lb{afe}
S =- \int d^4x\   [ \ti{B} (d\ti{A}  +
\fr{1}{2}\ti{A}\ti{A})  -\fr{1}{2} \ti{A}\ti{A}],
\ee
which yields
\beqa
S & = & -\int d^4x\ \{ B_{\mu\nu}F^{\mu\nu} +
2\ep_{\mu \nu \rho \sig }C_0^\mu D^\nu B^{\star \rho \sig}
+2C_1 D^\mu C^\star_{0\mu}  \nonumber \\
& & +\ep^{\mu \nu \rho \sig }
C_1 [B^\star_{\mu \nu }, B_{\rho \sig}^\star ] -
\fr{1}{2}A_\mu A^\mu \} ,  \lb{ath}
\eeqa
in terms of the components.
This is the minimal solution of the
master equation of the theory defined by
(\ref{si})\cite{af}, which can also be deduced by observing
that the transformations
\be
\lb{az}
\delta \ti{A}  = \ti{F}   ,\   \delta \ti{B}  = -\ti{D} \ti{B}  +\ti{A} ,
\ee
found by substituting (\ref{afe}) into (\ref{bt}), satisfy
\[
\begin{array}{lll}
\delta^2 \ti{A}   & = &  \ti{D} \cdot  \ti{F}=0 , \\
\delta^2\ti{B}  & = & -\ti{D}  \cdot
(- \ti{D} \ti{B} +\ti{A} ) - \ti{F} \ti{B}  + \ti{F}=0       ,
\end{array}
\]
due to the Bianchi identities and the definition
of the curvature.

\subsubsection{Chern-Simons theory in $d=2p+1$}

By examining the reducibility properties
(\ref{zrmc}) of the theory given by (\ref{cs})-(\ref{sl}),
one introduces ghosts,
ghosts of  ghosts and the related antifields which lead to
the generalized fields
\beqa
\ti{\phi}  &  =  &
\sum_{i=0}^{p-1}  \left[ \phi_{(2i+1,0)}
+ \sum_{j=1}^{2i+1} \eta_{(2i+1-j,j)}  +\phi^*_{(2i+2,-1)}
+ \sum_{j=-2p+2i+4}^{-2} \eta^*_{(2i+1-j,j)} \right]  \nonumber \\
 & &  \lb{ft} \\
\ti{\psi} &  =  &
\sum_{i=0}^p  \psi_{(2i,0)}
+\sum_{i=1}^p \sum_{j=1}^{2i} \kappa_{(2i-j,j)}
+\sum_{i=0}^p \psi^*_{(2i+1,-1)}
+ \sum_{i=0}^{p-1} \sum_{j=-2p+2i+1}^{-2} \kappa^*_{(2i-j,j)}  .\nonumber
\eeqa
The antifield of the field $a_{(k,l)}$ is defined as $a^*_{(2p+1-k,-l-1)}$.
Observe that  $\ti{\phi}$ and $\ti{\psi}$ are,
respectively,  collection of $2i+1$-forms
and  $2i$-forms. Now, in terms of
$\ti{A} =\ti{\phi}+\ti{\psi}$, we can write
\be
\lb{sd}
S_d=\fr{1}{2}\int  _{M_d} \left( \ti{A} d \ti{A}
+ \fr{2}{3} \ti{A}^3 \right) .
\ee
In terms of $\ti{\phi}$ and $\ti{\psi}$ components
(\ref{sd}) yields
\be
\lb{fa}
S_d = \int  _{M_d} \left( \ti{\phi} d \ti{\phi} +\fr{1}{3}\ti{\phi}^3
+\ti{\psi}(d+\ti{\phi} )\ti{\psi} \right) .
\ee

$S_d$ is the proper solution of the master equation,
because it is invariant under the transformations generated by
\[
\ti{R}=d+[ \ti{A}, ]=\fr{\del^2 S_d}{\del \ti{A}^2},
\]
due to the generalization of (\ref{rgt}).
Because of the sign assignments in (\ref{ft}) the transformations
(\ref{ggi}) are given as
\[
\D_{\ti{\Sigma}} \ti{A}_i=
\omega_{ij}\fr{\del_l \del_rS}{\del \ti{A}_j \del\ti{A}_k}
\ti{\Sigma }_k;\ \omega_{ij}=\left(
\ba{cc}
{\bf 0} & {\bf 1} \\
- {\bf 1} & {\bf 0}
\ea
\right) ,
\]
where the generalized gauge parameter is
\beqa
\ti{\Sigma} & = & \ti{\La} + \ti{\Xi}, \nonumber  \\
\ti{\La} &  =  &
\sum_{i=0}^{p-1}  \La_{(2i,0)}
+\sum_{i=1}^{p-1} \sum_{j=1}^{2i} \la_{(2i-j,j)}
+\sum_{i=0}^{p-1} \La^*_{(2i+1,-1)}
+ \sum_{i=0}^{p-2} \sum_{j=-2p+2i+1}^{-2} \la^*_{(2i-j,j)}  \nonumber \\
\ti{\Xi}  &  =  &
\sum_{i=0}^{p-1}  \left[ \Xi_{(2i+1,0)}
+ \sum_{j=1}^{2i+1} \xi_{(2i+1-j,j)}  +\Xi^*_{(2i+2,-1)}
+ \sum_{j=-2p+2i+4}^{-2} \xi^*_{(2i+1-j,j)} \right] . \nonumber
\eeqa
Observe that the transformations given by (\ref{bt}) by making
use of (\ref{sd}), leads to
\[
\delta \ti{A} =\ti{F} ,
\]
so that
\[
\delta^2 \ti{A} =0,
\]
following from the Bianchi identities.

This example is somehow different from the general case, because
the total degree of the components of $A$ are not the same.
But the integral selects only the terms with the correct degree.
One could write the solution of the master equation by using the
generalized forms each of which possessing only one degree,
and then gather them to obtain (\ref{sd}).

\subsubsection{The Gauge Theory of Quadratic Lie Algebras}

Although the gauge generators satisfy an open algebra,
the gauge theory of the quadratic Lie algebra (\ref{l0})-(\ref{g2}),
is an irreducible theory. Thus, one needs to introduce
only one family of ghosts $\eta^a.$
The generalized fields are
\beqa
\ti{h} = h_{(1,0)} +\eta_{(0,1)}+\Phi^\star_{(2,-1)}, \\
\ti{\Phi}=-h^\star_{(1,-1)}-\eta^\star_{(2,-2)}+\Phi_{(0,0)}.
\eeqa
Now, by replacing the fields $h,\ \Phi$ with the generalized
ones $\ti{h},\ \ti{\Phi}$ in (\ref{l0})
one obtains
\be
\lb{qlas}
S=-\int d^2x\fr{1}{2}\{ \ti{\Phi}_a(d \ti{h}^a
+{f_{bc}}^a \ti{h}^b\ti{h}^c +
V_{bc}^{ad} \ti{\Phi}_d\ti{h}^b\ti{h}^c)+k_{ab}\ti{h}^a\ti{h}^b \}.
\ee
It is the solution of the master equation, because
$S$ is invariant under the gauge
transformation obtained as the generalization of
the original ones
(\ref{g1})-(\ref{g2}), which can be written as in (\ref{ggi}):
\beqa
\left(
\begin{array}{c}
\D_{\ti{\la}} \ti{h} \\
\D_{\ti{\la}} \ti{\Phi}
\end{array}
\right)     & = &
\left(
\begin{array}{cc}
d+f\ti{h} + 2V \ti{\Phi} \ti{h} & V \ti{h} \ti{h} \\
f\ti{\Phi}+V\ti{\Phi} +k &  d+f\ti{h} + 2V \ti{\Phi} \ti{h}
\end{array}
\right)
\left(
\begin{array}{c}
\ti{\la} \\
0
\end{array}
\right)  .   \nonumber
\eeqa

As we discussed in Section \ref{sgs}, this symmetry yields
\[
\fr{\del (S,S)}{\del \ti{h}}=0,
\]
but $(S,S)$ cannot depend only on $\ti{\Phi}$
because there is not any field in $\ti{\Phi}$ whose
total degree is $(0,1),$
so that
\[
(S,S)=0.
\]

(\ref{qlas}) in components yields
\beqa
S= & \int d^2x \{ {\cal L}  +h^{\star \mu}_a (\del_\mu \eta^a
+{f_{ba}}^c\Phi_c \eta^b +
2V_{bc}^{ad} \Phi_d h_\mu^b\eta^c) & \nonumber \\
 & +\Phi^{\star a} ({f_{ba}}^c\Phi_c\eta^b +
V_{ba}^{cd}\Phi_c\Phi_d\eta^b +k_{ba}\eta^b) & \nonumber \\
& +\eta^\star_a (\fr{1}{2}{f_{bc}}^a \eta^b\eta^c +
V_{bc}^{ad}\Phi_d\eta^b\eta^c )
-\fr{1}{2} \ep_{\mu \nu} V_{bc}^{ad}
h^{\star \mu}_ah^{\star \nu}_d\eta^b\eta^c\} . & \lb{gsm}
\eeqa
Indeed, (\ref{gsm}) is the proper solution
of the master equation
as one can check explicitly.

\subsection{Discussions}

One can apply the formalism used in this paper to
other gauge systems like topological quantum field
theories\cite{nedall},\cite{ow3}
(for a review of topological field theories see \cite{tft})
and covariant string field theories.
In fact, without realizing the general formalism,
it was shown in the quantization of the Neveu-West
covariant string field theory\cite{nw} that the generalized fields
of Section \ref{sgs} can be used to write the proper solution
of the master equation\cite{omqstf}.

Although, gauge fixing can be performed in a compact way
in terms of generalized fields, particular properties
of the gauge system considered are essential to discuss it
in a concrete (not formal) manner. Because
of not being involved with a specific gauge theory,
here we discussed the gauge fixing on general grounds
without applying it to the examples considered.

\pagebreak

\newcommand{\bpl}{{\it Phys. Lett. }}
\newcommand{\mpl}{{\it Mod. Phys. Lett. }}
\newcommand{\np}{{\it Nucl. Phys. }}
\newcommand{\phr}{{\it Phys. Rep. }}
\newcommand{\dpre}{{\it Phys. Rev. }}

\end{document}